\pdfoutput=1
\documentclass{aa}  

\usepackage{graphicx}
\usepackage{txfonts}
\usepackage{hyperref}
\usepackage{natbib}
\usepackage{subcaption}

\usepackage{mwe}
\usepackage[labelfont=bf]{caption}
\usepackage{etex}
\reserveinserts{28}

\begin{document}

   \title{Metallicity Ceiling in Quasars from Recycled Stellar Winds}

   \author{Shelley J. Cheng
          \inst{1}
          \and
          Abraham Loeb\inst{1}\fnmsep
          }

   \institute{Harvard-Smithsonian Center for Astrophysics (CfA), Harvard University, 60 Garden Street, Cambridge, MA 02138, USA\\\email{shelley.cheng@cfa.harvard.edu}}

   \date{\today}

 
  \abstract
   {Optically luminous quasars are metal rich across all redshifts. Surprisingly, there is no significant trend in the broad-line region (BLR) metallicity with different star formation rates (SFR) and the average N V/ C IV metallicity does not appear to exceed $9.5~Z_\odot$. Combined, these observations may suggest a metallicity ceiling.}
   {Here, we conduct an exploratory study on scenarios relating to the evolution of embedded stars that may lead to a metallicity ceiling in quasar disks.}
   {We develop a simple model that starts with gas in a ``closed box'', which is enriched by cycles of stellar evolution until eventually newly formed stars may undergo significant mass loss before they reach the supernovae stage and further enrichment is halted. Using the \texttt{MESA} code, we create a grid over a parameter space of masses ($>8~M_\odot$) and metallicities ($1-10~Z_\odot$), and locate portions of the parameter space where mass loss via winds occurs on a timescale shorter than the lifetime of the stars. }
   {We find that for reasonable assumptions about stellar winds, sufficiently massive ($8-22~M_\odot$) and metal-rich ($\sim9~Z_\odot$) stars lose significant mass via winds and may no longer evolve to the supernovae stage, thereby failing to enrich and increase the metallicity of their surroundings. This suggests that a metallicity ceiling is the final state of a closed-box system of gas and stars.}
   {}


   \maketitle
%

\section{Introduction} \label{sec:intro}

In the unified model of Active Galactic Nuclei (AGN) \citep[e.g.][]{Krolik:88, Antonucci:93, Netzer:15, Elvis:20}, a central black hole is surrounded by an accretion disk and a dusty thick torus that may act as an obscuring medium. Quasars are believed to be one of many observed phenomena that can be explained by AGNs, with Type I and Type II quasars differentiated, respectively, by broad and narrow spectral lines. According to the unified model of AGN, Type I quasars are due to a more face-on viewing angle that has a line-of-sight towards the faster-moving Doppler-broadened emission lines from the inner broad-line region (BLR). Type II quasars are believed to be AGN that are viewed from a more edge-on angle such that the dusty torus obscures the BLR. 

While the unified model of AGN is powerful in its ability to explain a variety of phenomena such as quasars, blazars, and Seyfert galaxies, an independent model that appeals to changing AGN accretion rates complements the unified model \citep[e.g.][]{Nicastro:00, Elitzur:09, Trump:11, Best:12}. When considering accretion rates (rather than only viewing angle), quasars are typically classified as embedded or optically luminous. Embedded quasars are thought to feature a geometrically thick accretion disk with low accretion rate that prevents the formation of broad emission lines. With the accretion rate in the disk low, the gas and dust form stars with a high star formation rate (SFR). Alternatively, an optically luminous quasar features a higher accretion rate in a geometrically thin accretion disk, with a lower SFR since the gas and dust are consumed in the accretion stream. It is thought that embedded quasars (higher SFR) evolve to become optically luminous quasars (lower SFR) when accretion clears the surrounding gas and dust \citep[e.g.][]{Murray:05, Rieke:88, Izumi:18}. Therefore, the SFR is expected to change with time as the quasar transitions from embedded to optically luminous.

Metallicity is often used as an important probe of past star formation, since cycles of star formation progressively enrich the interstellar medium (ISM) with metal-rich material. Therefore, it is interesting that despite the expectation for SFR to change over time as quasars evolve from embedded to optically luminous, quasars are metal rich across all redshifts \citep{simon:10,shemmer:04,warner:04,nagao:06,jiang:07,juarez:09}. Previous studies regarding this puzzle have concluded that broad-line region (BLR) metallicity is independent of the ongoing star formation of the host galaxy, which is responsible for the far-infrared (FIR)-bright quasars. Instead, the BLR metallicity is correlated to enrichment from an earlier stage of star formation \citep{simon:10}. Also, interestingly, there is no significant trend in the BLR metallicity with SFR, showing an average N V/ C IV metallicity not exceeding $9.5$ solar metallicities \citep{simon:10}. Additional observational work involving composite AGN spectra by \citet{warner:03, warner:04, Mignoli:19} have measured N V/ C IV line ratios that agree with \citet{simon:10}, and correspond to average N V/ C IV metallicities not exceeding $\sim10$ solar metallicities using the correlation described in \citet{Hamann:2002} with updates to the solar abundance ratios by \citet{Dhanda:07}. We note that the conversion from N V/ C IV flux line ratios to metallicity involves photoionization model calculations, which can produce varying metallicity results depending on model details such as zoning, densities, and ionization degrees. In this work, the N V/ C IV metallicity results presented by \citet{simon:10} are taken to be a fiducial point of comparison. Incidentally, this N V/ C IV metallicity corresponds to $\sim 20~\%$ of the mass gas budget composed by heavy elements. Since changes in SFR are thought to occur over time in the aforementioned accretion rate model of AGNs (as they transition from embedded and optically luminous), this lack of trend in BLR metallicity with differing SFR may be suggestive of a metallicity ceiling.

Line-driven winds are driven by photon momentum transfer through metal line absorption and provides a source of mass loss for hot stars \citep{kudritzki:00}. Since these winds involve metal line absorption, it is not surprising that wind momenta and thus mass loss rate directly correlates with metallicity \citep{lucy:70,castor:75,vink:21}, with stars of higher metallicities experiencing greater mass loss. In some cases, this mass loss can be so great that stars lose their envelope and never burn heavier metal elements, ending their lives as Wolf-Rayet stars \citep{sander:20}. As cycles of stellar formation take place and the ISM is continually enriched, stars will eventually experience significant mass loss due to winds.

In this paper, we explore whether mass loss in stars due to winds can lead to a metallicity ceiling in a ``closed-box.'' We create a grid of models using the \texttt{MESA}\footnote{\href{https://docs.mesastar.org/en/release-r21.12.1/index.html}{https://docs.mesastar.org/en/release-r21.12.1/index.html}} code that span a range of masses and metallicities (Section~\ref{sec:methods}) and determine which systems fail to reach the supernova minimum mass due to winds, where the mass loss timescale is shorter than the stellar lifetime (Section~\ref{sec:results}). Additional tests are presented in Section~\ref{sec:extra} that show more details about the dependence of the results on wind model parameters and metallicity. Finally, we offer a discussion of our results before drawing our conclusions (Section~\ref{sec:discussion}).

\section{Methods} \label{sec:methods}
Since quasars feature deep gravitational potential wells and have short dynamical times, a ``closed box'' model that allows for many generations of star formation is a suitable first-order approximation. Our box extends across the BLR region for the inner $\sim1$~parsec, and is ``closed'' in the sense that the majority of freshly enriched material from cycles of star formation will remain within the box. While supernovae ejecta can reach velocities of $20,000~$km/s, the vast majority of the ejected mass of heavy elements is traveling below the BLR escape velocity of $\sim7000-10,000~$km/s \citep[e.g.][]{Miniutti:2014, Muller:2020, Ni:2018, Sun:2023} even for supernovae at high explosion energies; see for example \citet{Goldberg:19}. Additionally, the fastest moving supernovae ejecta are also the least metal-rich; see Figure 9 in \citet{Goldberg:19}. On the sub-pc scale, some ultra-fast outflows can be at velocities of up to $\sim20,000~$km/s (greater than the escape speed), but can only lead to mass loss of $\sim1~M_\odot/\text{year}$ from the BLR \citep[see review][]{Morganti:2017}. We note that the timescale between each generation of stars in our ``closed box'' is on the order of millions of years while the timescale for a quasar to remove enriched gas is on the order of hundreds of millions of years, so enrichment (and therefore metallicity) will saturate before star formation is suppressed. Therefore, even if enriched material is lost through outflows that exceed the BLR escape velocity, metallicity will nevertheless increase over time and eventually saturate over a timescale shorter than the quasar timescale. A more quantitative analysis that describes how factors such as accretion onto the black hole affect metallicity may require a full numerical simulation that goes beyond the scope of this work.

To explore the possibility of a metallicity ceiling in this ``closed box'' framework, it is sufficient to study the evolution of the last cycle of stars. We used version \texttt{r21.12.1} of the Modules for Experiments in Stellar Astrophysics code
(\texttt{MESA}) \citep{paxton:11,paxton:13,paxton:15,paxton:18,paxton:19} to evolve our stars. MESA is a one-dimensional stellar evolution code that numerically solves the fully coupled structure and composition stellar equations for stars of a wide range of masses and ages. MESA includes modules that provide equations of state, opacities, nuclear reaction rates, element diffusion data, and atmosphere boundary conditions. \footnote{The \texttt{MESA} equation of state (EOS) combines OPAL \citep{Rogers2002}, SCVH
\citep{Saumon1995}, FreeEOS \citep{Irwin2004}, HELM \citep{Timmes2000},
PC \citep{Potekhin2010}, and Skye \citep{Jermyn2021} EOSes. Radiative opacities combines OPAL \citep{Iglesias1993,
Iglesias1996} and data from \citet{Ferguson2005} and \citet{Poutanen2017}.  Electron conduction opacities are from
\citet{Cassisi2007}. Nuclear reaction rates are from JINA REACLIB \citep{Cyburt2010}, NACRE \citep{Angulo1999} and \citet{Fuller1985, Oda1994,
Langanke2000}.  Screening is included via the prescription of \citet{Chugunov2007}.
Thermal neutrino loss rates are from \citet{Itoh1996}. }

We focused on stars with masses $>8~M_\odot$, since only these massive stars can end their evolution in a Type II supernova and increase the metallicity of the surrounding gas (following canonical stellar evolution). We created a grid over a parameter space of varying masses ($8-22~M_\odot$) and metallicities ($3~Z_\odot$, $5~Z_\odot$ and $9~Z_\odot$), and located portions of the parameter space where mass loss via winds occurs on a timescale shorter than the lifetime of stars. These regions represent configurations where stars will fail to enrich the surrounding gas with heavy elements, leading to a metallicity ceiling in the surroundings.

The different metallicity choices of $3~Z_\odot$, $5~Z_\odot$ and $9~Z_\odot$ were implemented in \texttt{MESA} by altering the \texttt{initial\_z} and \texttt{Zbase} parameters, with chosen \texttt{Zbase} values of $0.04$, $0.06$, and $0.12$ for the respective \texttt{initial\_z} metallicity choices. We approximate the lifetime of stars as the time at which core Helium burning begins using the analytic expressions for $t_{\text{He}}$ presented in \cite{hurley:2000}.

For winds, we adopted the \texttt{Dutch} wind model, which follows \cite{Glebbeek:2009} in combining the models of \cite{Vink:2001} and \cite{Nugis:2000}. The \texttt{Dutch} wind model is sophisticated and depends on a variety of parameters including metallicity, stellar mass, luminosity, temperature, terminal velocity, and Helium mass fraction. The strength of the \texttt{Dutch} wind, controlled by the \texttt{Dutch\_scaling\_factor}, is not definitively agreed upon in literature, with studies suggesting mass loss rates that differ from those predicted by \texttt{Dutch} wind based on arguments related to observations or numerical treatment \citep[e.g.][]{Belczynski:22,Agrawal:22,Crowther:10}. For instance, a lower value of \texttt{Dutch\_scaling\_factor} of $0.8$ has been used in \citet{Fields:18, Perna:18}, \texttt{Dutch\_scaling\_factor}$=1.0$ has been used in \citet{Goldberg:20, Higgins:20}, and higher \texttt{Dutch\_scaling\_factor} values of $1.5$, $1.9$ and $2.0$ have been featured in \citet{Quataert:16, Belczynski:22}. In our exploratory study, we heuristically treat \texttt{Dutch\_scaling\_factor} as a variable parameter and set the \texttt{Dutch\_scaling\_factor} to $2.0$ for our initial study. The effect of a reduced \texttt{Dutch\_scaling\_factor} is described in Section~\ref{sec:extra_wind}. Higher values of \texttt{Dutch\_scaling\_factor} beyond $2.0$ have not been considered in the literature for stellar evolution except in the context of stripping stars for use as supernovae progenitors (e.g. see \citet{Shiode:14}), and is therefore not relevant to this work. Additionally, mass loss due to super-Eddington luminosity was included for completeness, with a \texttt{super\_eddington\_wind\_Ledd\_factor} of $1.0$. The inlists used for the models were adapted from the \texttt{high\_z} models available in the \texttt{MESA} test suites.

\begin{figure*}

    \centering
\begin{subfigure}[b]{0.49\textwidth}
	\includegraphics[width=1\textwidth]{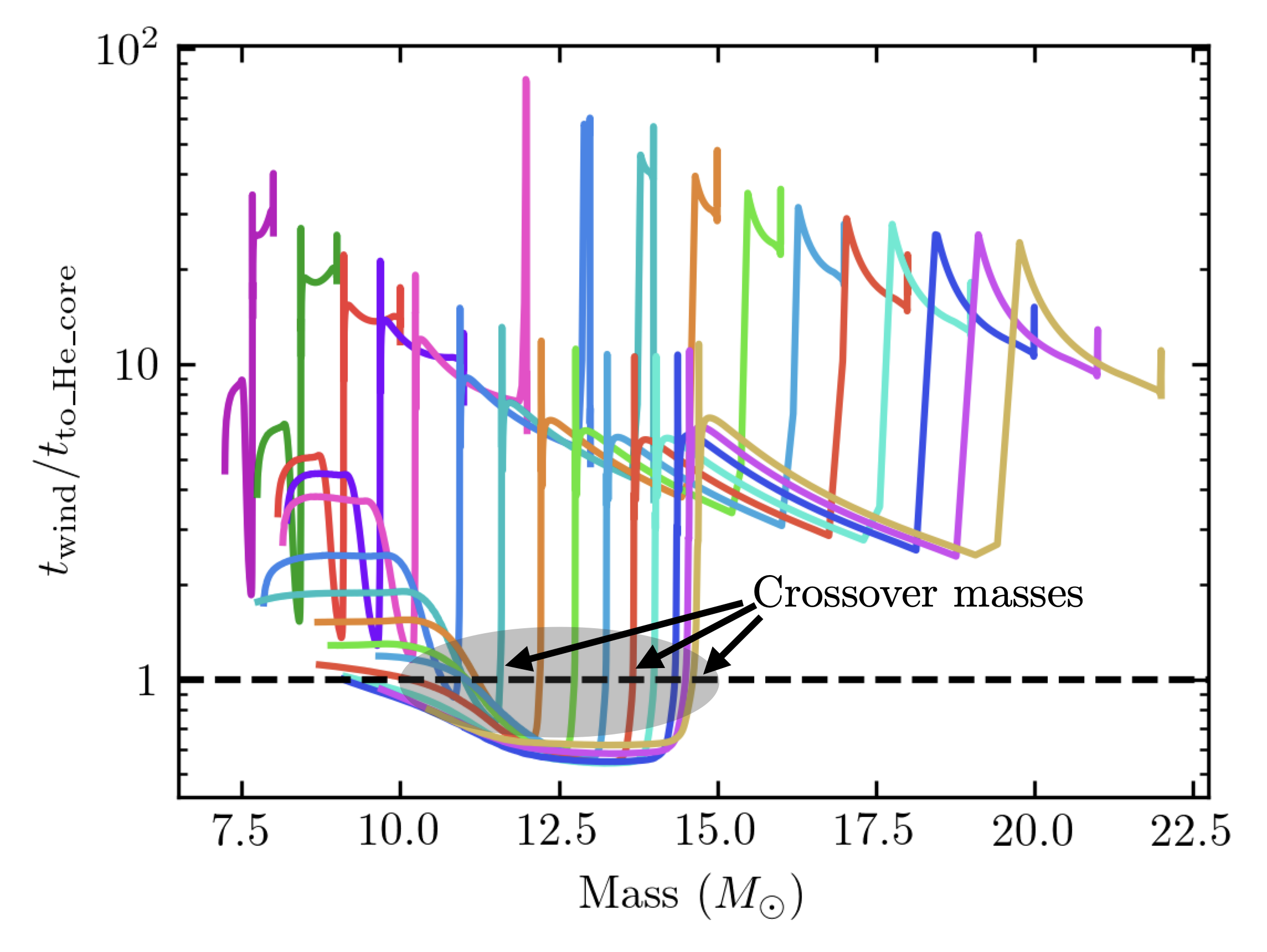}
	\caption{}
\end{subfigure}
\begin{subfigure}[b]{0.49\textwidth}
	\includegraphics[width=1\textwidth]{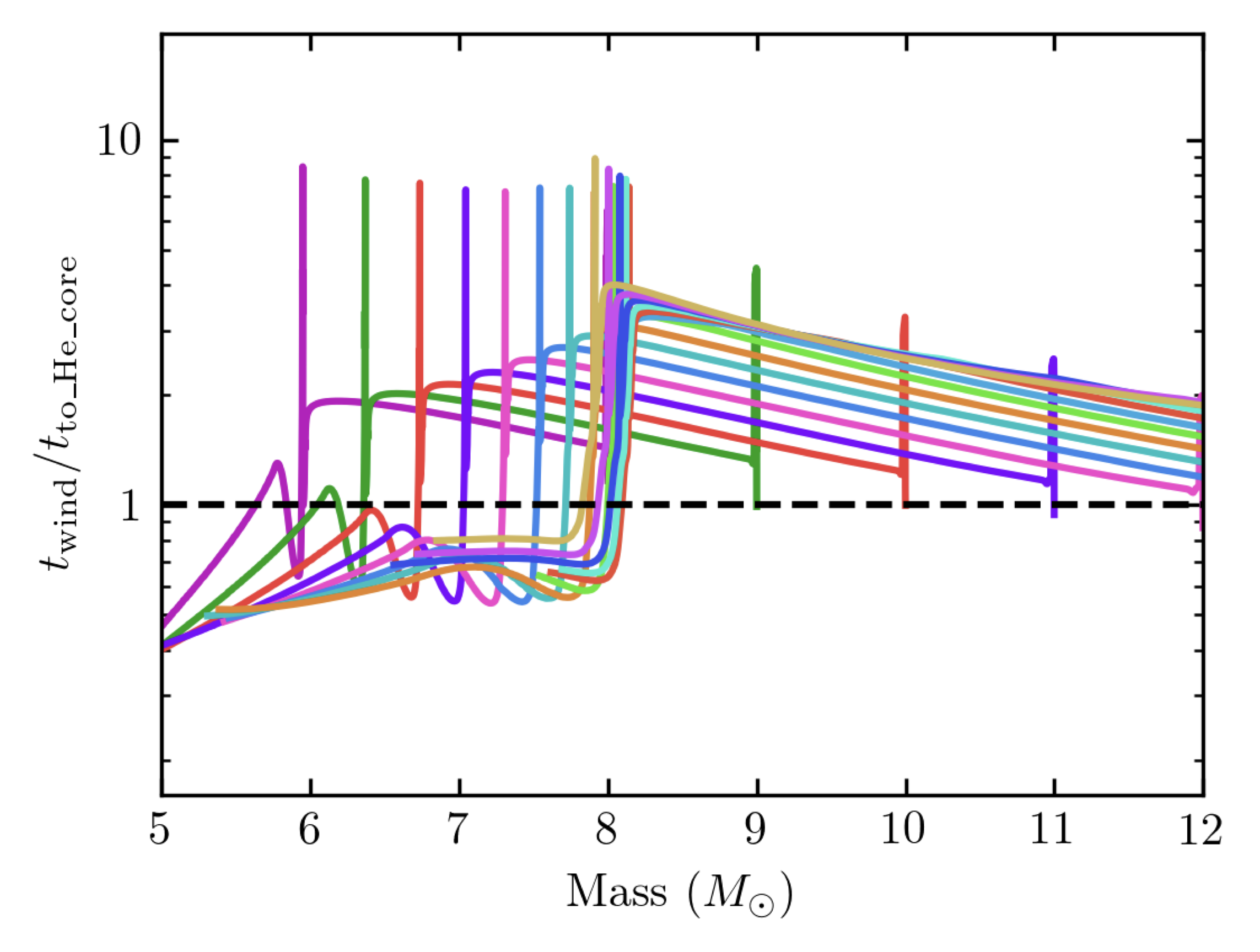}
	\caption{}
\end{subfigure}

\centering
\hspace{-100pt}
\begin{subfigure}[b]{0.49\textwidth}
	\includegraphics[width=1.5\textwidth]{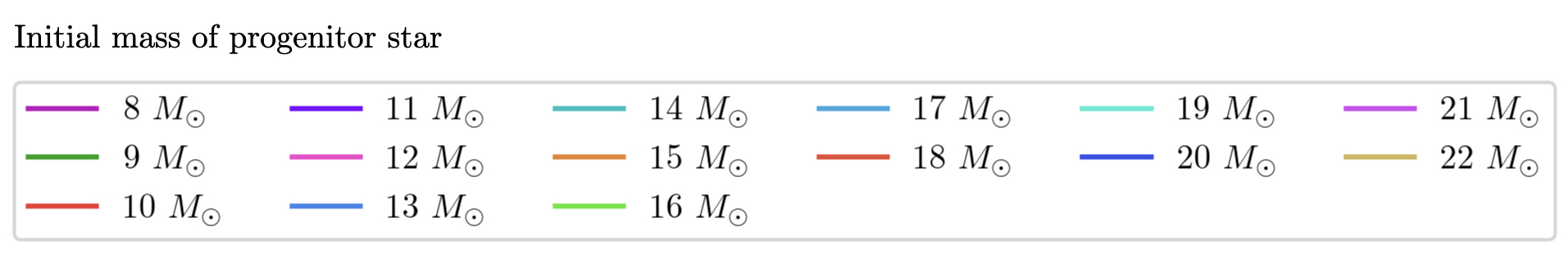}
\end{subfigure}

\caption{Ratio between mass-loss timescale and lifetime of stars modeled with \texttt{MESA}. Panel (a): Metallicity $=5~Z_\odot$. Panel (b): Metallicity $=9~Z_\odot$ The horizontal axis shows the total stellar mass, and the vertical axis shows the timescale ratio (see text for more details). The colored tracks show the evolution of the mass loss timescale for different initial mass (see legend). Stellar evolution starts on the right of the plot and progresses towards the left as mass is lost. The black dashed line is displayed for clarity and shows when the wind-lifetime ratio equals to $1$. At a metallicity of $5~Z_\odot$ (left panel), only systems with $M>13~M_\odot$ have a wind-lifetime ratio that falls below unity, with varying crossover masses. The panel on the right, at $9~Z_\odot$, shows that all modeled systems have wind-lifetime ratios that crosses below $1$. The gray shaded oval and the black arrows in the left panel represents the crossover of the wind-lifetime ratio below unity and defines the crossover mass (see Section~\ref{sec:results} for more details).}
\label{fig:wind}
\end{figure*}

\section{Results} \label{sec:results}
We determine whether mass loss due to winds is sufficient to significantly impact stellar evolution by plotting the ratio between the wind mass loss timescale, $t_{\text{wind}}$, and the lifetime left until the start of Helium core burning, $t_{\text{to\_He\_core}}$. $t_{\text{to\_He\_core}}$ decreases in value as the star evolves; {$t_{\text{to\_He\_core}} = t_{\text{He}} - t_i$ where $t_i$ is the age of a star at time $i$. $t_{\text{wind}}$ is defined as

\begin{equation}
t_{\text{wind}} = \frac{M}{\dot{M}_{\text{wind}}} 
\end{equation}

with the mass of the star as $M$ and the mass loss rate due to winds as $\dot{M}_{\text{wind}}$.

The full result is shown in Figure~\ref{fig:wind}. For clarity, an example evolutionary track is shown in Figure~\ref{fig:evol} with the stages labeled. When this wind-lifetime timescale ratio is less than $1$ ($t_{\text{wind}} < t_{\text{to\_He\_core}}$), the mass loss due to winds is significant enough to potentially prevent supernova (if the star's mass falls below $\sim8~M_\odot$).

As seen in Figure~\ref{fig:wind} (a), models with $5~Z_\odot$ require initial stellar masses of $M>13~M_\odot$ for the wind-lifetime timescale ratio to dip below $1$. Conversely, as shown in Figure~\ref{fig:wind} (b), the wind-lifetime timescale ratio of models with $9~Z_\odot$ across all modeled masses ($8-22~M_\odot$) falls below $1$ as the stars transition from Hydrogen core burning to Helium burning. This indicates that mass loss due to winds is significant for systems at very high metallicities.

\begin{figure*}
    \centering
    \begin{subfigure}[b]{0.49\textwidth}
        \includegraphics[width=\textwidth]{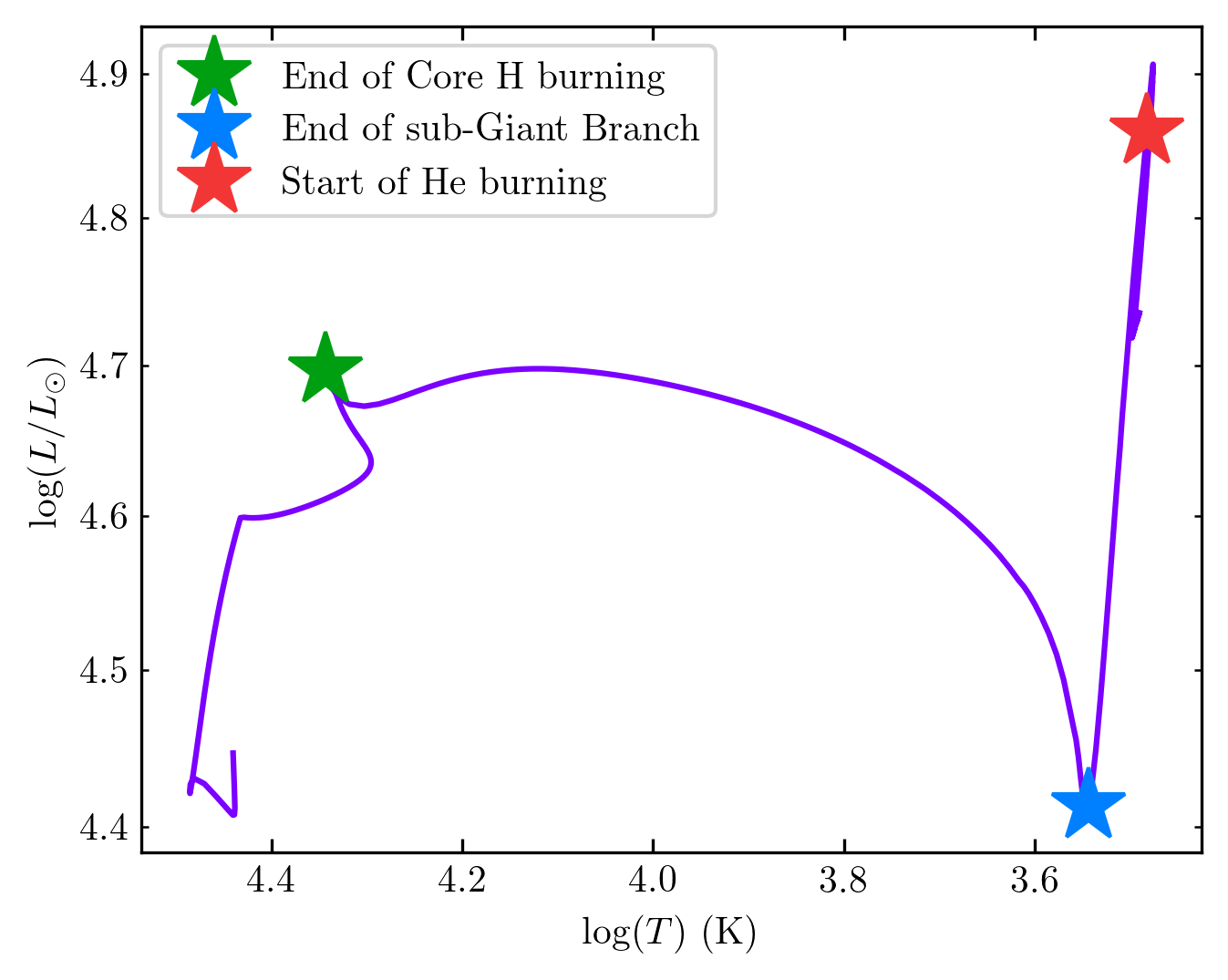}
        \caption{}
    \end{subfigure}
    \begin{subfigure}[b]{0.49\textwidth}
        \includegraphics[width=\textwidth]{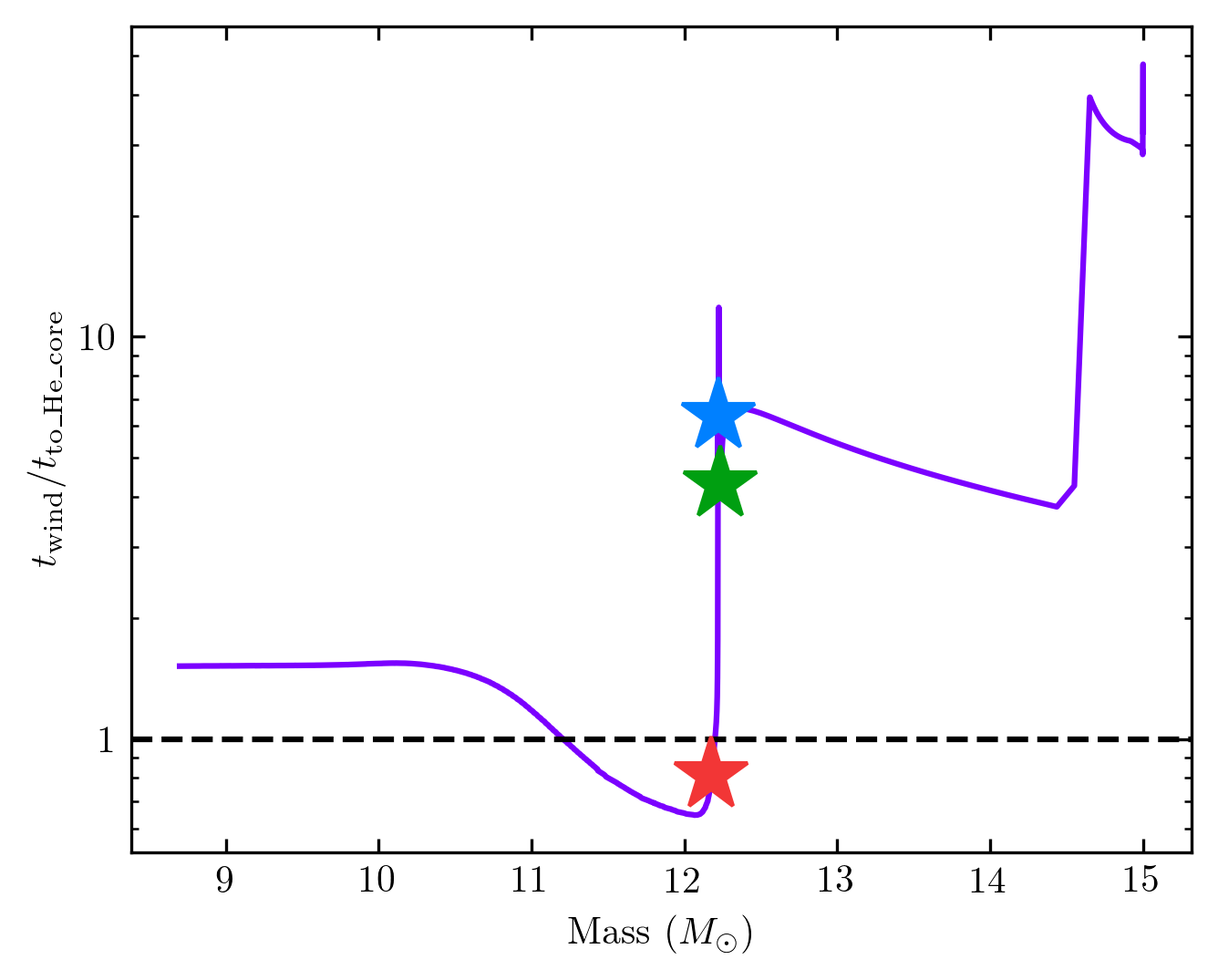}
        \caption{}
    \end{subfigure}
    \caption{Evolutionary stages of a $15~M_\odot$ star with  $5~Z_\odot$ modeled with \texttt{MESA}. Panel (a): H-R Diagram, with key evolutionary points marked with colored stars. Panel (b): Wind-lifetime ratio vs. mass (one of the tracks from Figure~\ref{fig:wind}), with the same evolutionary points marked. For this example system (and indeed all other systems), the steepest wind-lifetime ratio decrease occurs when the star halts core H burning and begins He burning. We note that the He burning shown here (marked with the red star) represents the first instance of a decrease in total Helium mass, and precedes the core He burning defined in \cite{hurley:2000}.}\label{fig:evol}
\end{figure*}

\begin{figure}[ht!]
	\includegraphics[width=\columnwidth]{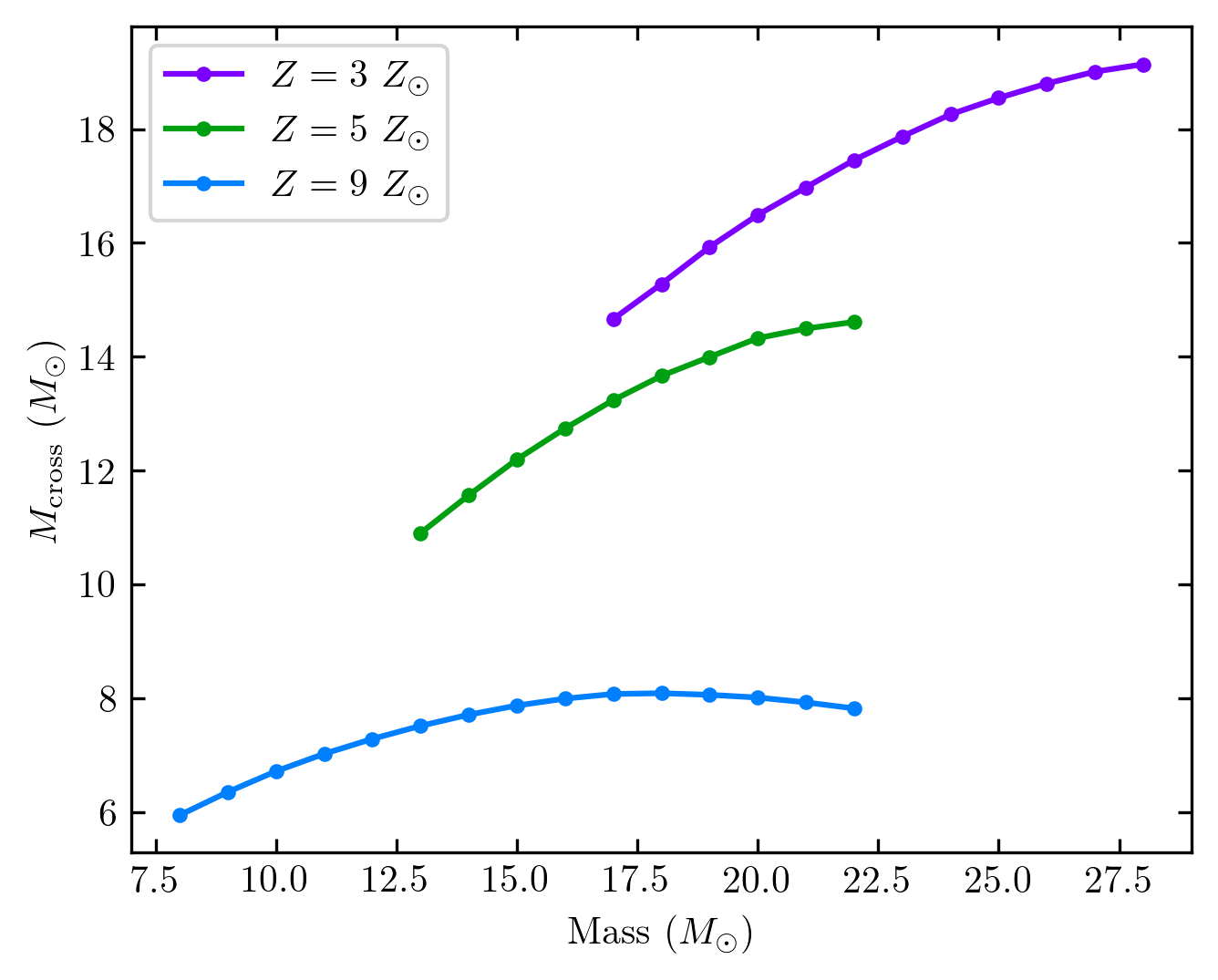}
	\caption{Crossover mass of all modeled systems. The horizontal axis shows the initial mass of the star, and the vertical axis shows the crossover mass, namely the mass at which the mass loss time scale is shorter than the remaining stellar lifetime (see Figure~\ref{fig:wind}). Lower mass stars than shown for $Z=3~Z_\odot$ or $Z=5~Z_\odot$ do not have a crossover mass (see Figure~\ref{fig:wind} for $Z=5~Z_\odot$ example). Stars with $Z=3~Z_\odot$ or $Z=5~Z_\odot$ clearly have $M_{\text{cross}} \gtrsim 8~M_\odot$, while stars with $Z=9~Z_\odot$  have $M_{\text{cross}} \lesssim 8~M_\odot$. This implies that stars forming in a $Z=9~Z_\odot$ ISM will fail to undergo Type II supernovae and will lead to a metallicity ceiling in the ISM. }
	\label{fig:crossover}
\end{figure}

The mass at which the timescale ratio equals $1$ is defined as the crossover mass $M_{\text{cross}}$. If $M_{\text{cross}}\lesssim8~M_\odot$, then the system loses sufficient mass and will no longer end its life in a Type II supernova and would therefore fail to further enrich the ISM. We adopted this notion of a crossover mass due to the limitations of \texttt{MESA} in modelling massive stars at high metallicities at more advanced evolutionary stages.

We emphasize that the existence of a crossover mass alone is not sufficient to make conclusions about whether the system undergoes a Type II supernova. The mass \textit{value} at which crossover occurs must be below $8~M_\odot$ for the system to no longer undergo a supernova. Thus, the crossover mass value acts as a simple test of whether a star undergoing wind-driven mass loss will lose enough mass to avoid a supernova.

Figure~\ref{fig:crossover} shows the crossover mass of all models where the timescale ratio reached below $1$, for all modeled metallicities. We clearly see that $3~Z_\odot$ and $5~Z_\odot$ systems have $M_{\text{cross}}>8~M_\odot$, and Type II supernovae and enrichment can continue to occur. Therefore metallicities below $5~Z_\odot$ are insufficient to produce a metallicity ceiling.

However, for $9~Z_\odot$ systems, we see that $M_{\text{cross}}\lesssim8~M_\odot$ for all modeled masses with a decreasing trend towards the higher masses. This indicates that, at very high metallicities of around $9~Z_\odot$, massive stars with $M>8~M_\odot$ fail to retain enough mass to end their lives as Type II supernovae. Therefore, our model heuristically demonstrates that stars will no longer undergo supernova and cannot further enrich the ISM once the ISM metallicity reaches around $9~Z_\odot$. A metallicity ceiling is thus possible once stars begin to form in a $9~Z_\odot$ ISM. Taking into account the intrinsic uncertainties of modelling massive stars at high metallicities with wind-driven mass-loss, this result is consistent with the observed average BLR metallicity of no more than $9.5~Z_\odot$ \citep{simon:10}.

\section{Additional Tests} \label{sec:extra}
\subsection{Metallicity} \label{sec:extra_metallicity}
Figure~\ref{fig:metals_explicit} explicitly shows the effect of different metallicities on the evolution and crossover mass of a $17~M_\odot$ star (with $\texttt{Dutch\_scaling\_factor}=2$). Other than metallicity, all other model parameters were identical between the $3$ models shown in Figure~\ref{fig:metals_explicit}. It is clear that higher $Z$ leads to more efficient winds and greater mass loss before the onset of helium burning in the stars (refer to Figure~\ref{fig:evol} for evolutionary stages). This can be explained by the direct dependence on $Z$ of the \texttt{Dutch} wind model as well as the effect of $Z$ on other stellar parameters relevant in the wind model (refer to Section~\ref{sec:methods} for wind model references and further details). The crossover masses likewise decrease with increasing $Z$. 

\begin{figure}[ht!]
	\includegraphics[width=\columnwidth]{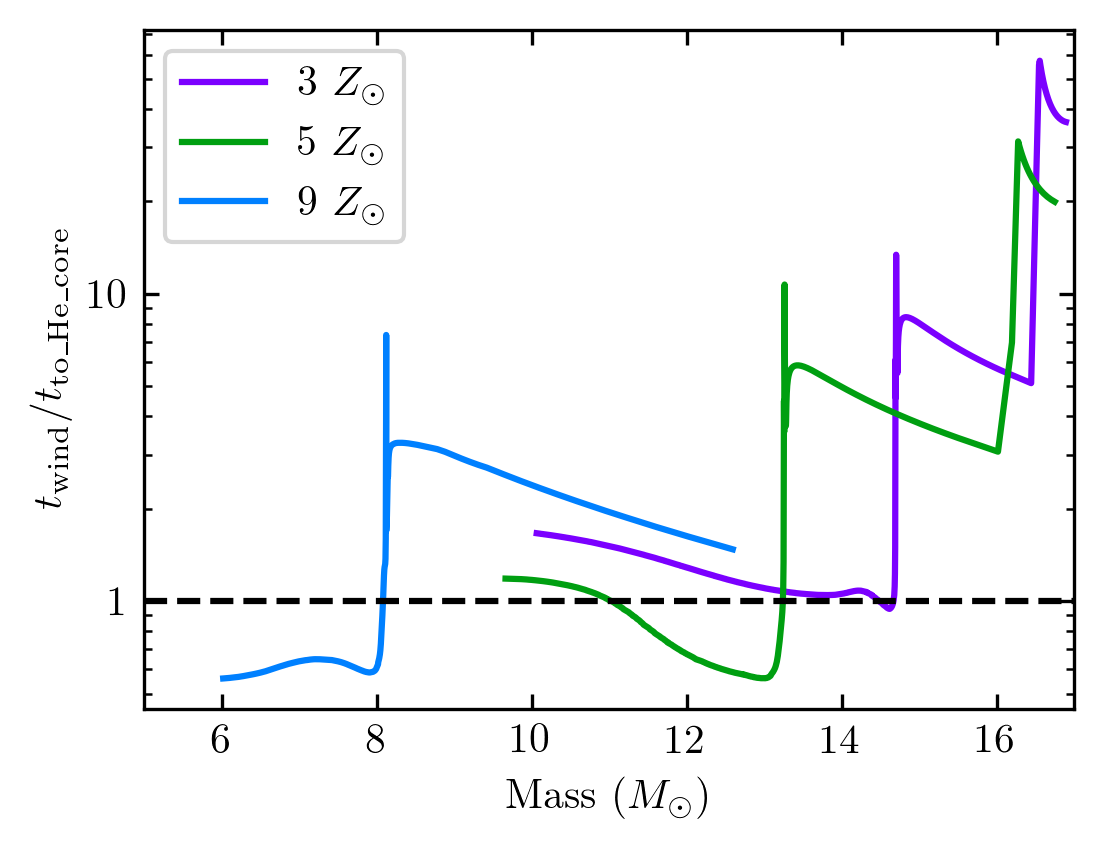}
	\caption{Wind-lifetime ratio vs. mass of a $17~M_\odot$ star at different metallicities. The axes are the same as those in Figure~\ref{fig:wind} and the right panel of Figure~\ref{fig:evol}. This plot combines the $17~M_\odot$ tracks previously shown in Figure~\ref{fig:wind} (at $5~Z_\odot$ and $9~Z_\odot$) with the $3~Z_\odot$ model. The crossover masses decrease with increasing metallicity (see text). The first $2~$million years of evolution has been excluded for plot clarity. }\label{fig:metals_explicit}
\end{figure}

\begin{figure}[ht!]
	\includegraphics[width=\columnwidth]{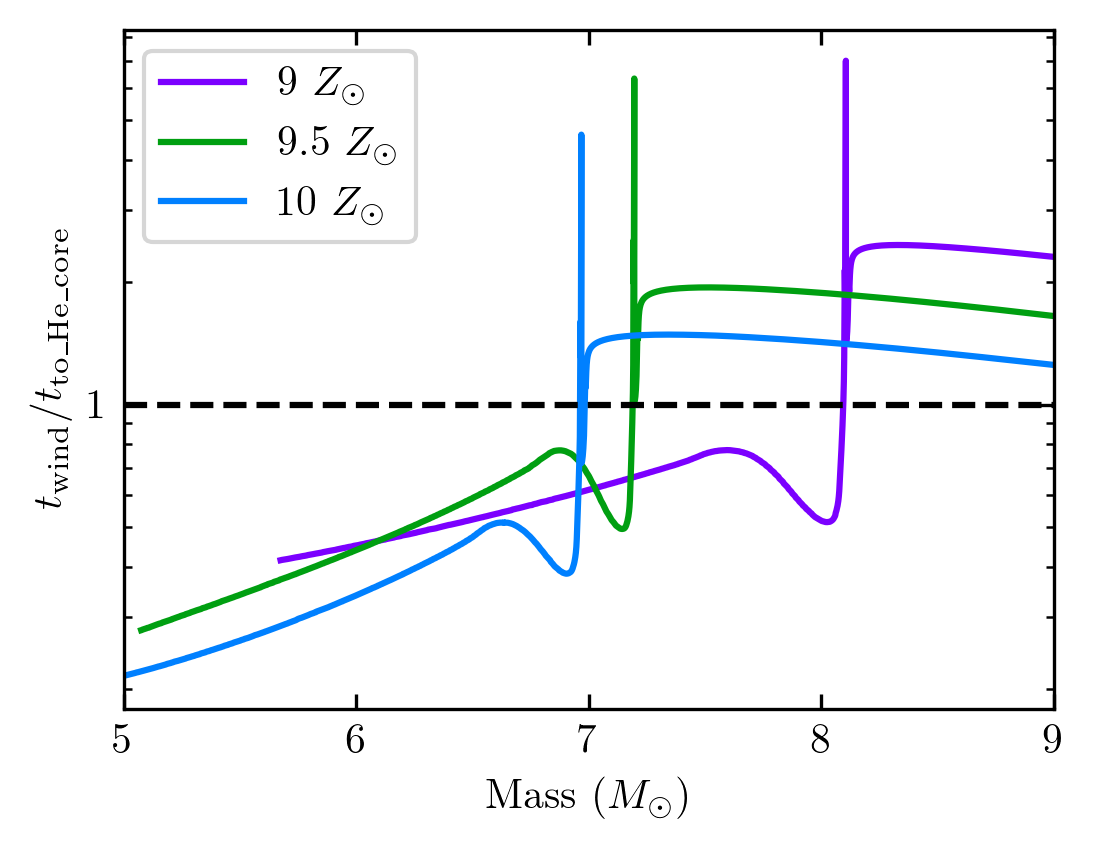}
	\caption{Wind-lifetime ratio vs. mass of a $12~M_\odot$ star with reduced winds at different metallicities. The axes are the same as those in Figure~\ref{fig:wind} and the right panel of Figure~\ref{fig:evol}. The crossover masses decrease with increasing metallicity (see text).}\label{fig:lesswind}
\end{figure}

\subsection{Wind Model} \label{sec:extra_wind}
As mentioned in Section~\ref{sec:methods}, we selected a scaling factor of $2$ for the \texttt{Dutch} wind model. However, since the modeling of massive star evolution at high metallicities using \texttt{MESA} is overall uncertain, we conducted a test with a $25\%$ reduction in the \texttt{Dutch} wind scaling factor on a $12~M_\odot$ star at $9~Z_\odot$. 

When \texttt{Dutch\_scaling\_factor} was set to $2$, the crossover mass of the system was $7.3~M_\odot$ (see the blue line in Figure~\ref{fig:crossover}). Since this crossover mass is below the minimum supernovae mass of $8~M_\odot$, the system would not undergo supernova in its lifetime.

However, when this \texttt{Dutch\_scaling\_factor} was reduced by $25\%$, the crossover mass for this $12~M_\odot$ star at $9~Z_\odot$ increased to $8.1~M_\odot$ and therefore the star would undergo supernova. At first glance, this would seem to suggest that our results are sensitive to the choice of \texttt{Dutch\_scaling\_factor}. However, this is not the case, since a small increase in metallicity would decrease the crossover mass and compensate for the effect of a lower wind scaling factor. As shown in Figure~\ref{fig:lesswind}, an increase of metallicity by $0.5~Z_\odot$ to a value of $9.5~Z_\odot$ is sufficient to decrease the crossover mass down to $7.2~M_\odot$ (see green line in Figure~\ref{fig:lesswind}). Since the possible metallicity ceiling from \cite{simon:10} has sizable uncertainties, this $0.5~Z_\odot$ change in metallicity is acceptable. Therefore, the results of this work remain qualitatively valid with changes in the \texttt{Dutch} wind scaling factor.

\section{Discussion and Conclusions} \label{sec:discussion}
We investigated whether stars embedded near quasars (quasar stars) can lead to a metallicity ceiling based on simple closed box model. We assumed that the stars evolve over many cycles and progressively increase the metallicity within the closed system, and explicitly studied the last cycle of stars that may remain massive enough to evolve and enrich the ISM via Type II supernovae. Observationally, evidence of a metallicity ceiling in quasars was presented in \cite{simon:10} who found that the broad-line region metallicity average at no more than $9.5~Z_\odot$ with no significant dependence on star formation rate (SFR).

We evolved stars of masses $8-22~M_\odot$ and $3~Z_\odot$, $5~Z_\odot$ and $9~Z_\odot$ using \texttt{MESA} to explore the mass-metallicity parameter space required for a metallicity ceiling. We found that, in an existing ISM metallicity of $9~Z_\odot$, stars with masses $8-22~M_\odot$ lose enough mass before core Helium burning to fall below the $\sim8~M_\odot$ Type II supernova threshold, and will therefore lead to a metallicity ceiling. We determined that lower ISM metallicities of $5~Z_\odot$ and $3~Z_\odot$ are insufficient for a metallicity ceiling, and therefore conclude that a minimum ISM metallicity of $9~Z_\odot$ is needed for saturation of the enrichment process. Consequently, a metallicity ceiling may be a general phenomenon in environments that can be approximated as stellar formation and evolution within a closed-box. 

The idea of a metallicity ceiling in quasars has implications to our understanding of black hole accretion and nuclear star clusters (NSCs). For instance, stellar formation at the outer regions of quasar disk \citep{goodman:04} may be a possible explanation for periods of lower-than-expected black hole accretion rates, since the gas that otherwise would be accreted instead fragments and form stars (and accretion is choked). However, as the metallicity of stars become too high to maintain stability after many cycles, gas may no longer be trapped in stars, and thus black hole accretion rates increase to expected levels (i.e. choked accretion is lifted). If we adopt the metallicity ceiling idea for star formation in the outer regions of the disk, changes in the black hole accretion rates may reveal information about the fraction of inflowing gas that reaches the inner accretion disk versus the fraction that remains in the outer star-forming region.

Additionally, there is an almost-constant ratio between the mass of NSCs and the mass of the central black hole \citep{neumayer:20} in observations. If NSCs form with the presence of a metallicity ceiling that lifts choked accretion, this almost-constant ratio of black hole mass and NSC mass may be explained. 

The metallicity ceiling discussed in this paper is still valid in binary and multiple stellar systems. Though most massive stars are in binary or multiple-star systems
\citep[e.g.][]{Sana:2012}, binary interactions would only serve to increase metallicity through mixing, leading to the system reaching a metallicity ceiling quicker. 
As the accretor star gain mass, it experiences stronger winds, loses mass more rapidly than it gains mass (in \texttt{MESA}'s \texttt{Dutch} wind model), and therefore will more likely fall below $\sim8~M_\odot$. Likewise, the donor star will lose mass and more likely fall below $\sim8~M_\odot$. Therefore, the presence of multiple-star systems would only lead to the system suppressing supernova more effectively. 

Additionally, we can interpret our result through the lens of stellar masses. When supernovae are suppressed, a mass cut-off past the minimum supernova mass of $\sim8~M_\odot$ is expected. Therefore, the presence of a highly supersolar metallicity ceiling of $\sim 10~Z_\odot$ suggests a mass cut-off above $\sim8~M_\odot$. 

We emphasize that the metallicity ceiling and suppression of supernova described in this work only applies to environments that can be approximated as a closed-box. For more realistic environments, in-flowing fresh gas will be able to maintain $Z<8~Z_\odot$ and allow supernova, and a metallicity ceiling will not be present.

\section*{Acknowledgments}
We thank Evan Bauer for advice and comments on \texttt{MESA} wind modeling, and Matteo Cantiello, Charlie Conroy, Lars Hernquist, and Mark Reid for helpful comments on the manuscript. 

\bibliography{aanda}{}
\bibliographystyle{aa}

\end{document}